\documentclass[aps,pra,amsfonts,superscriptaddress,twocolumn]{revtex4} 
\usepackage{epsfig}
\usepackage{bbm} 
\usepackage{multirow}
\usepackage{lscape}
\usepackage{rotating,placeins}
\usepackage{hyperref}
\newcommand{\eq}{\begin{eqnarray}} 
\newcommand{\en}{\end{eqnarray}}
\def\bra#1{\mathinner{\langle{#1}|}}
\def\ket#1{\mathinner{|{#1}\rangle}}

\newcommand{\nicefrac}[2]{{#1}/{#2}}
\usepackage{hyperref}
\usepackage{amsmath} 
\usepackage{placeins}

\begin{document}

\title{Many-particle interference beyond  many-boson and many-fermion statistics} 
\author{Malte C. Tichy}
\affiliation{Physikalisches Institut, Albert--Ludwigs--Universit\"at Freiburg, Hermann--Herder--Strasse~3, D--79104 Freiburg, Germany}
\affiliation{Lundbeck Foundation Theoretical Center for Quantum System Research, Department of Physics and Astronomy, University of Aarhus, DK--8000 Aarhus C, Denmark}
\author{Markus Tiersch}
\affiliation{Physikalisches Institut, Albert--Ludwigs--Universit\"at Freiburg, Hermann--Herder--Strasse~3, D--79104 Freiburg, Germany}
\affiliation{Institute for Quantum Optics and Quantum Information, Austrian Academy of Sciences, Technikerstrasse 21A, A--6020 Innsbruck, Austria}
\author{Florian Mintert}
\affiliation{Physikalisches Institut, Albert--Ludwigs--Universit\"at Freiburg, Hermann--Herder--Strasse~3, D--79104 Freiburg, Germany}
\affiliation{Freiburg Institute for Advanced Studies, Albert-Ludwigs-Universit\"at, Albertstrasse 19, 79104 Freiburg, Germany}
\author{Andreas Buchleitner}  
\affiliation{Physikalisches Institut, Albert--Ludwigs--Universit\"at Freiburg, Hermann--Herder--Strasse~3, D--79104 Freiburg, Germany}

\date{\today}

\begin{abstract}Identical particles exhibit correlations even in the absence of inter-particle interaction, due to the exchange (anti)symmetry of the many-particle wavefunction. Two fermions obey the Pauli principle and anti-bunch, whereas two bosons favor bunched, doubly occupied states.  Here, we show that the collective interference of three or more particles leads to a much more diverse behavior than expected from the boson-fermion dichotomy known from quantum statistical mechanics. The emerging complexity of many-particle interference is tamed by a simple law for the strict suppression of events in the Bell multiport beam splitter. The law shows that counting events are governed by widely species-independent interference, such that bosons and fermions can even exhibit identical interference signatures, while their statistical character remains subordinate. Recent progress in the preparation of tailored many-particle states of bosonic and fermionic atoms promises experimental verification and applications in novel many-particle interferometers.
\end{abstract}
\maketitle

\section{Introduction}
The symmetrization postulate enforces the (anti)symmetrization of the bosonic (fermionic) many-particle wavefunction \cite{Girardeau:1965ys} and  thereby severely restricts the set of accessible states for indistinguishable particles. When one postulates that each microscopic state is populated with equal probability \cite{Romer:1994uq}, the resulting statistical physics of bosons, fermions and distinguishable particles heavily differs, as directly observed in two-point correlation functions. The latter reveal bunching of bosons \cite{Hanbury-Brown:1956vn,Hodgman25022011} and the opposed anti-bunching of fermions \cite{Henny:1999cr,Kiesel:2002dq,Rom:2006uq}, which can also be directly compared in a single setup \cite{Jeltes:2007ly}. The differences between the species is often ascribed a rather universal character and said to be rooted in many-particle interference \cite{Rom:2006uq,Jeltes:2007ly,Folling:2005fk}. However, the states that are prepared in such many-body experiments are thermal, such that only the \emph{statistical} behavior of bosons and fermions is probed, and no coherent many-particle interference crystallizes out, as we will explain further down. 

As a prominent example for such many-particle interference, and in only ostensible agreement with the \emph{statistical} behavior of \emph{many} bosons, two single photons exhibit the Hong-Ou-Mandel \cite{Hong:1987mz} effect: Two indistinguishable photons that fall simultaneously onto the input modes of a beam splitter with reflectivity 1/2 always bunch and leave the setup together. Fermions behave in the opposite way, the Pauli principle enforces them to anti-bunch and to choose distinct output modes \cite{PhysRevA.58.4904}. Also for more than two photons, bosonic effects  boost the probability to find all particles in one output mode \cite{Ou:2006ta,Ou:1999lo,Ou:1999rr,Niu:2009pr}. Many-particle interferences thus seem to boil down to a behavior that is familiar from statistical physics.

Here we falsify this popular view and show that scattering events with many particles and many modes give rise to much richer many-particle interference phenomena than intuitively expected from few-particle interference and quantum statistical mechanics. The behavior of bunched bosons and anti-bunched fermions, which was also presumed to dominate many-particle interference \cite{Lim:2005qt}, is widely insufficient for the understanding of the coherent behavior of many particles, since interference obtrudes the overall picture. In particular, bosonic and fermionic interference effects are not necessarily opposed to each other, but fermions can experience fully destructive interference in the very same way bosons do. Only when the purity of the initial state is destroyed, many-particle interference patterns vanish, and a smooth, familiar, bosonic or fermionic behavior is recovered. 

We establish our results by theoretically comparing the scattering of distinguishable particles to bosons and fermions in a setup with many modes. Particles are initially prepared in $n$ input modes, they then scatter off a common potential, such that each particle ends in a coherent superposition of the $n$ output modes. The probability for a counting event, {\it i.e.}~for an event with a certain number of particles in each output mode, then reflects many-particle interference of bosons and fermions, when contrasted to distinguishable particles, for which usual combinatorial laws apply. Our system of $n$ input- and $n$ output modes can be realized with multiport beam splitters for photons \cite{Reck:1994zp} or by an appropriate sequence of tunneling couplings in experiments with ultracold atoms in optical lattices \cite{Weitenberg:2011vn,Tichy:2011vn}. The distinguishability of the particles can be achieved by misaligning the path lengths for photons, or by populating different internal hyperfine states for cold atoms. Given the recent breakthroughs in the control and measurement of single optically trapped atoms \cite{Sherson:2010fk,Weitenberg:2011vn,Bakr:2010fk,Karski:2009vn}, the experimental verification of the \emph{coherent} collective behavior of many particles, which we will show to heavily contrast the familiar established statistical effects, is in reach. 

\section{Framework}
\subsection*{Single-particle evolution}
For our focus on many-particle interference, we exclude any interparticle interaction. The many-particle behavior can then be inferred from the single-particle time-evolution by appropriately summing the emerging \emph{many-particle} amplitudes or probabilities. Under the action of the single-particle time-evolution operator $\hat U$, an input state $\ket{\phi^a_j}$ evolves into a superposition of output states $\ket{\phi^b_k}$,  
\eq \hat U \ket{\phi^a_j} &=& \sum_{k=1}^n U_{j,k} \ket{\phi^b_k}  \label{unitary1q} ,
 \en 
 where $a$ ($b$) refers to input (output) modes, and $j$ ($k$) is the respective mode number; the first (second) index of any scattering matrix refers to the input (output) mode. The probability 
  \eq p_{j,k} = |U_{j,k}|^2, \label{singleparticleprob}  \en for a particle in the $j$th input mode to reach the $k$th output mode contains all occurring double-slit like single-particle interference. For identical particles, it is useful to work in second quantization: The creation operators $\hat a^\dagger_j$ and $\hat b^\dagger_j$ for particles in the input- and output-modes, respectively, inherit the relationship (\ref{unitary1q}),
\eq \hat a^\dagger_j \rightarrow \sum_{k=1}^n U_{j,k} \hat b^\dagger_k , \label{timeevocreation} \en
which is valid for, both, fermions and bosons. The behavior of \emph{many-particle} states is governed by the (anti)commutation relations for the creation and annihilation operators for (fermions) bosons. 

\subsection*{Particle arrangements}
Initially, $N$ particles are prepared in the $n$ input modes, with  $r_j$ particles in the $j$-th mode, {\it i.e.}~$\sum_{j=1}^n r_j=N$. We denote such \emph{arrangement} of particles in the modes by a unique input \emph{mode occupation list} 
$\vec r=(r_1, \dots , r_n)$ of length $n$. 
For distinguishable particles, the specification of the particle arrangement leaves the freedom to distribute the (labeled) particles among the input modes in several different ways. Without restriction of generality, we choose the initial state 
\eq \ket{\Psi^{\text{in}}}_{\text{D}} = \otimes_{j=1}^n \left( \otimes_{k=1}^{r_j} \ket{\phi^a_j} \right) , \en
in the first-quantization formalism. For bosons and fermions, the specification of a mode occupation list $\vec r$ specifies the initial quantum state, 
\eq \ket{\Psi^{\text{in}}}_{\text{F}/\text{B}} = \prod_{j=1}^n \frac{ \left( \hat a^\dagger_j \right)^{r_j} }{\sqrt{r_j!} } \ket{0} ,\en
in the second-quantization formalism, where $\ket{0}$ denotes the vacuum. 

After time-evolution according to (\ref{unitary1q}) and (\ref{timeevocreation}), respectively, the number of particles in each output mode is measured. Events are characterized by the corresponding particle arrangement defined by the \emph{output} mode occupation list $\vec s=(s_1,s_2,\dots,s_n)$, where again $\sum_{j=1}^n s_j =N$.

\begin{figure*}[ht] \center
\includegraphics[width=.8\linewidth,angle=0]{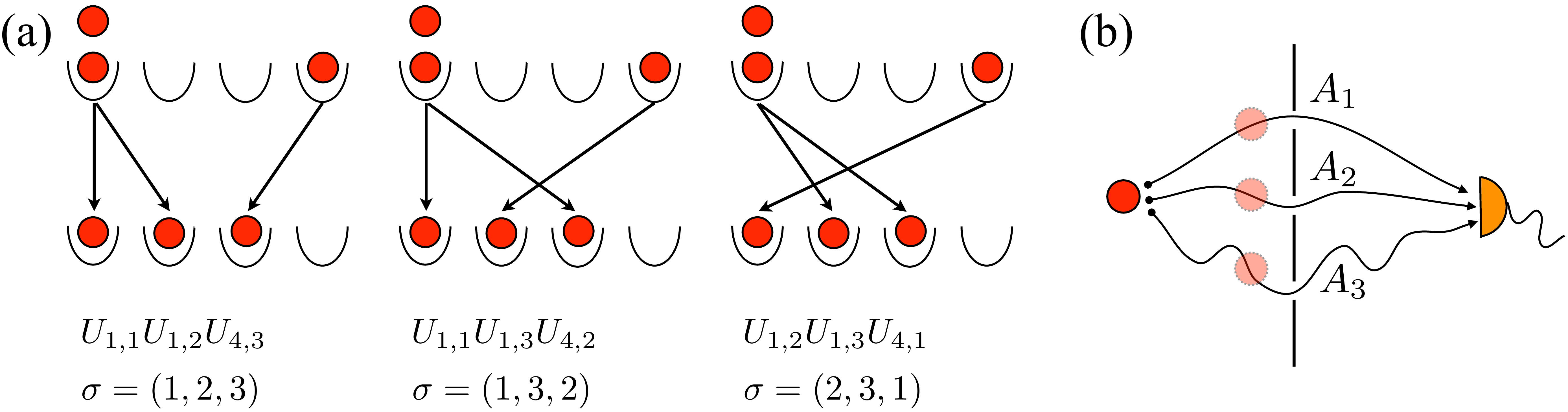} 
\caption{{\bf Many-particle paths and single-particle analogy.} (a) The mode occupation list of the initial (upper) arrangement reads $\vec r=(2,0,0,1)$, the equivalent mode assignment list is $\vec d(\vec r)=(1,1,4)$, see Eq.~(\ref{drepres}). The final arrangement is characterized by $\vec s=(1,1,1,0)$ and $\vec d(\vec s)=(1,2,3)$. Three physically distinct many-particle paths, each corresponding to a permutation $\sigma$ of $\vec d(\vec s)$, connect the arrangements. The many-particle (path) amplitudes are the corresponding products of single-particle amplitudes, given here below the respective paths. They are summed for all possible permutations in Eqs.~(\ref{distpermanents}) and (\ref{bosontranspermurep}) to give the full transition probability and amplitude, respectively. (b) Single-particle analogy: A single particle can pass through three slits, which corresponds to three distinct complex amplitudes. Again, the initial and final state are connected through three single-particle paths.  }\label{pathsfig} \end{figure*}

It is convenient to define for each (input or output) arrangement $\vec q$ an alternative notation $\vec d(\vec q)$, the \emph{mode assignment list}. The list is of length $N$, with entries that specify each particle's provenience or destination. It is constructed by concatenating $q_j$ times the mode number $j$:
\eq \vec d(\vec q) =\oplus_{j=1}^n \oplus_{k=1}^{s_j} (j) = (\underbrace{1,\dots,1}_{q_1},\underbrace{2,\dots,2}_{q_2}, \dots, \underbrace{n,\dots,n}_{q_n}) \label{drepres} \en  
The relationship between mode occupation- and mode assignment lists is exemplified in Fig.~\ref{pathsfig}.

Particle arrangements with cyclic symmetry will play an important role in our subsequent treatment. For any integer $m$ that divides $n$, we define an arrangement $\vec q$ of $N$ particles to be $m$-\emph{periodic} if it consists of $p=\nicefrac n m$ repetitions of a pattern $\vec k$ of length $m$. The mode occupation list thus reads 
\eq \vec q=( \underbrace{
\underbrace{k_1, k_2, \dots, k_{m}}, \underbrace {k_1, \dots, k_{m}}, \dots, \underbrace {k_1, \dots, k_{m}}}_{p=\nicefrac n m} ) , \label{defiperiodicstate} \en
while the mode assignment list $\vec d(\vec q)$ satisfies
\eq \forall j: d_{j+\nicefrac N p}(\vec q)= d_{j}(\vec q)+m ~, \label{propper} \en
where we identify $d_{N+j} \equiv d_{j}$ and $d_j \equiv  d_j+n$. For example, the arrangement $\vec q=(2,1,0,5,0,2)$ exhibits no periodicity, and $m=n=6, p=1$; the strongest symmetry is exhibited by an arrangement with $\forall j: q_j=r_1$, as, {\it e.g.}, $ \vec q=(4,4,4,4,4)$, for which $p=n, m=1$. 

\subsection*{Transition probabilities}
Signatures of many-particle interference appear in the transition probability, \eq P_{\text{D/B/F}}(\vec r, \vec s;  U)  ,  \label{theprob} \en for an input arrangement $\vec r$, to an output arrangement $\vec s$, given the single-particle evolution $U$. Bosons (B) and fermions (F) can thus be compared  to distinguishable particles (D). 

For the latter, the probability to find the final arrangement $\vec s$ is obtained combinatorially by taking into account all possibilities to distribute the particles among the output modes, given the single-particle probabilities $p_{j,k}$ in Eq.~(\ref{singleparticleprob}): 
\eq P_{\text{D}} (\vec r, \vec s; U) &=& \sum_{\sigma \in S_{\vec d(\vec s)} }   \prod_{j=1}^N p_{d_{j}(\vec r),\sigma(j)}  ,  \label{distpermanents}  \en
where $S_{\vec d(\vec s)}$ denotes the permutations of the output mode assignment list $\vec d(\vec s)$. We define the $N\times N$ matrix \eq M_{j,k}=U_{d_j(\vec r),d_k(\vec s)} , \label{definM} \en such that 
\eq 
P_{\text{D}} (\vec r, \vec s; U) &=& \frac{1}{\prod_{j=1}^n s_j!} \text{perm}(| M|^2) , \label{distpermdd}
\en
where the absolute-square is understood to be taken component-wise and $\text{perm}(|M|^2)$ denotes the \emph{permanent} of $|M|^2$ \cite{Ryser:1963oa,Sachkov:2002fk}.  Each summand in (\ref{distpermanents}) represents one way to distribute the particles among the output modes with  $s_k$ particles in the $k$th mode. Each possibility corresponds to a \emph{many-particle path} as illustrated in Fig.~\ref{pathsfig}(a). 

For \emph{indistinguishable} particles, all many-particle paths contribute \emph{coherently} to the final state, and their \emph{amplitudes} need to be summed, such that 
\eq 
P_{\text{B}/\text{F}}(\vec r, \vec s ; U) &=&  \frac{\prod_j s_j!}{\prod_j r_j!}  \times \label{generalampli} \\ &&   \left| \sum_{\sigma \in S_{\vec d(\vec s)}} \text{sgn}_{\text{B}/\text{F}}(\sigma) \prod_{j=1}^N U_{d_j(\vec r),\sigma(j)} \right|^2 , \nonumber 
 \en 
 where $\text{sgn}_{\text{B}}(\sigma)=1$ and $\text{sgn}_{\text{F}}(\sigma)=\text{sgn}(\sigma)$ allows for the fermionic anti-commutation-relation. The transition amplitudes for bosons (fermions) can be re-written as the permanent (determinant) of $M$: 
\eq
P_{\text{B}}(\vec r, \vec s; U) &=& \frac{1}{\prod_{j} r_j! s_j!} \left| \text{perm}(M) \right|^2  \label{bosontranspermurep} , \\ 
P_{\text{F}}(\vec r, \vec s; U ) &=& | \text{det} (M) |^2 \label{determinantferm} ,
\en
in close analogy to (\ref{distpermdd}).

\subsection*{Single-particle analogy}
An analogy between coherent and incoherent \emph{many}-particle processes with the corresponding single-particle phenomena is suggestive: Consider a single particle that can pass through $N!$ distinct slits to fall onto a chosen point on a screen, as sketched in Fig.~\ref{pathsfig}. If the time evolution is coherent, {\it i.e.}~if the path taken by the particle is not observed,  the amplitudes of the $N!$ paths are summed, just like in Eq.~(\ref{generalampli}). The probability to observe the particle reflects changes of the relative phases of the superimposed path amplitudes.  Since the phases that are accumulated between each slit and the observation point can be, in principle, adjusted independently, fully destructive or strongly constructive interference can be induced. If the time-evolution, instead, occurs in an incoherent way, because,~{\it e.g.},~the path information is leaked to the environment, the respective path \emph{probabilities} need to be added, and the dependence on the accumulated phases vanishes, similar to Eq.~(\ref{distpermanents}).

In the many-particle situation, we deal with a total of 
\eq N_{\text{arr}}^{\text{D}}=N_{\text{arr}}^{\text{B}}={ N+n -1 \choose n-1}=\frac{(N+n-1)!}{(n-1)! N!} \label{numarrbos} \en distinct particle arrangements for bosons or distinguishable particles, and  
\eq N_{\text{arr}}^{\text{F}} ={n \choose N}=\frac{n!}{N! (n-N)!} , \label{numarrbosf} \en fermionic arrangements, {\it i.e.}~the number of possible counting events rapidly grows with the number of particles and modes. Again, the variation of the phases that are accumulated by the particles can be monitored by the  many-particle event probabilities $P_{\text{B/F}}$ given in (\ref{generalampli}). In contrast to the single-particle scenario, the up to $N!$ different many-particle amplitudes that enter in (\ref{generalampli}) are \emph{not} independently adjustable, since they are all given by products of single-particle matrix elements $U_{j,k}$. Therefore, the behavior of many-particle scattering systems of moderate size ($N\approx n \approx 10$) already presents a challenge, which is aggravated by the large number of events (\ref{numarrbos}) and (\ref{numarrbosf}). 

The aforementioned scaling argument also immediately exposes the main challenge for the exploitation of many-particle interference: 
 For fermions,  the Pauli principle implies $N \le n$, and $M$, given in (\ref{definM}), is a submatrix of $U$. Transitions are thus suppressed when $M$ is singular, according to Eq.~(\ref{determinantferm}) that relates the transition amplitude to the \emph{determinant} of $M$.  For bosons, the \emph{permanent} of $M$ governs the behavior, for which no analogous criterion for its vanishing exists \cite{Minc:1984uq}. For distinguishable particles, we face the permanent of a positive matrix in Eq.~(\ref{distpermdd}), which can be approximated efficiently \cite{Jerrum:2004:PAA:1008731.1008738}. The design of a setup that exhibits strong and controlled many-particle interference is thus in general much more involved than in the single-particle case.

\section{Suppression law}
A systematic assessment of many-particle interference becomes possible by imposing symmetries on the scattering setup. We therefore focus on the Bell multiport beam splitter \cite{PhysRevA.55.2564}, which is described by an \emph{unbiased} scattering matrix in the sense that all single-particle probabilities are equal, $|U_{k,l}|^2=p_{k,l}=1/n$. The scattering matrix describes a discrete Fourier transformation,  
\eq \left( U_n^{\text{Fou}} \right)_{j,k}= \frac{1}{\sqrt n} e^{i \frac{2 \pi}{n} (j-1)(k-1)} \label{foumadef} .\en
The phase that a particle acquires thus depends on the input and on the output mode, and can only assume a multiple of $\pi /n$.  In this setup, we expect a large visibility in the sense that fermionic/bosonic probabilities strongly differ from their counterparts with distinguishable particles, since many amplitudes of equal modulus, but of different phase, are added in Eq.~(\ref{generalampli}).  For distinguishable particles, Eq.~(\ref{distpermanents}) gives a purely combinatorial expression:
\eq P_{\text{D}}(\vec r, \vec s; U^{\text{Fou}}) = \frac{N!}{n^N \prod_{j=1}^n s_j!} ,  \label{distsimple} \en
{\it i.e.}~events occur with probabilities according to a \emph{multinomial distribution} \cite{Ryser:1963oa}, which generalizes the binomial distribution  found for a two-mode beam-splitter \cite{springerlink:10.1007}. 

For bosons and fermions, the evaluation of the transition amplitude (\ref{generalampli}) is, in general, not significantly simpler for the highly symmetric matrix (\ref{foumadef}) than for the general case \cite{Graham:1976nx}. As a result of $\left(U^{\text{Fou}}\right)^{-1}=\left( U^{\text{Fou}}\right)^*$, we can, however, exploit an input-output symmetry: \eq P_{\text{B/F}}(\vec r,\vec s; U^{\text{Fou}}_n) = P_{\text{B/F}}(\vec s, \vec r; U^{\text{Fou}}_n) \label{nonsymm} . \en
Additionally, arrangements that can be related to each other by a cyclic permutation or by reversing the mode order are equivalent, {\it i.e.}~the application of these transformations on the initial or final state does not change any event probability. 

Combining the symmetry properties of the Fourier matrix (\ref{foumadef}) for $m$-periodic ({\it i.e.}~cyclicly symmetric) initial or final arrangements, as introduced in Eqs.~(\ref{defiperiodicstate}), (\ref{propper}),  we can formulate a sufficient criterion for the occurrence of fully destructive interference and thus for the strict suppression of a transition $\vec r \leftrightarrow \vec s$, which very considerably generalizes a results for $N$ bosons that are prepared in $\vec r_c=(1,1,\dots, 1)$ and transmitted through an $N$-port beam splitter \cite{Lim:2005qt,Tichy:2010kx}. 

For bosons, given an $m$-periodic initial state $\vec r$, final states $\vec s$ are suppressed when the sum of the elements of their mode assignment list, $\vec d(\vec s)$, multiplied by the period length of the initial state, $m$, is not dividable by $n$, {\it i.e.~}
\begin{widetext}
\eq \text{mod}\left(m  \sum_{j=1}^N d_j(\vec s),n \right) ~ \neq~ 0 \ \Rightarrow \ P_{\text{B}}(\vec r, \vec s; U^{\text{Fou}}) =0 \label{law} . \en

For fermions, the anti-commutation relation leads to a behavior that depends on the parity of the number of particles: 
\begin{subequations}
\label{lawfermi}
\eq 
 N\text{ odd, or }\nicefrac{N}{p} \text{ even:}\text{ mod}\left(m  \sum_{j=1}^N  d_j(\vec s),n \right) \neq 0 &\Rightarrow& P_{\text{F}}(\vec r, \vec s; U^{\text{Fou}})=0 \label{lawfermi1} ,\\ 
N\text{ even and }\nicefrac{N}{p} \text{ odd: }\text{mod}\left(m  \sum_{j=1}^N  d_j(\vec s),n \right) \neq \frac n 2 &\Rightarrow & P_{\text{F}}(\vec r, \vec s; U^{\text{Fou}})=0  \label{lawfermi2} ,
\en
\end{subequations}
\end{widetext}
where $n$ is even when $N$ is even and $N/p$ is odd. The proof for the suppression law is given in the Methods section. Intuitively speaking, Eqs.~(\ref{law}) and (\ref{lawfermi}) formalize the observation that amplitudes of equal modulus but different phase annihilate each other when they are distributed in an equally spaced way on a circle around the origin in the complex plane.

The suppression laws (\ref{law}), (\ref{lawfermi}) circumvent the inherent complexity of expression (\ref{generalampli}): Even for very large particle numbers, the suppression of an event can be predicted easily, whereas the computation of the permanent (\ref{bosontranspermurep}) is computationally hard and thus practically unfeasible for combinations of large $N$ and $n$. For fermions, the determinant (\ref{determinantferm}) is computationally less expensive than the permanent, but the suppression law (\ref{lawfermi}) still offers a significant speedup: The evaluation of the determinant (\ref{determinantferm}) by the LU decomposition scales with $N^3$, while (\ref{lawfermi}) scales linearly in $N$. The suppression law also  encompasses several other criteria for the suppression of bosonic events in the literature as special cases \cite{nockdipm,Campos:1989fk,Lim:2005qt,springerlink:10.1007,Tichy:2010kx}. It does not give, however, a necessary criterion: It is possible to arrange amplitudes in the complex plane such that they cancel each other, while they do not lie on a circle. 

Formally, the suppression of events given a cyclically symmetric initial state can be interpreted as the manifestation of periodicity of the initial state in its Fourier transformation, given by Eq.~(\ref{foumadef}). Very distinct initial states that share the same periodicity exhibit the same strict suppression among the final states. This reflects their shared periodicity, whereas the information on the exact constitution of the arrangement is contained in the unsuppressed states. The single-particle transformation that corresponds to the Fourier matrix can also be seen as the transformation to the quasi-momentum basis \cite{al:2006kx}, and the quasi-momentum-distribution can be probed \cite{al:2006vn} to characterize the phase of a gas.

\subsection*{Scattering of bosons}
\begin{figure*} \center
\includegraphics[width=.8\linewidth,angle=0]{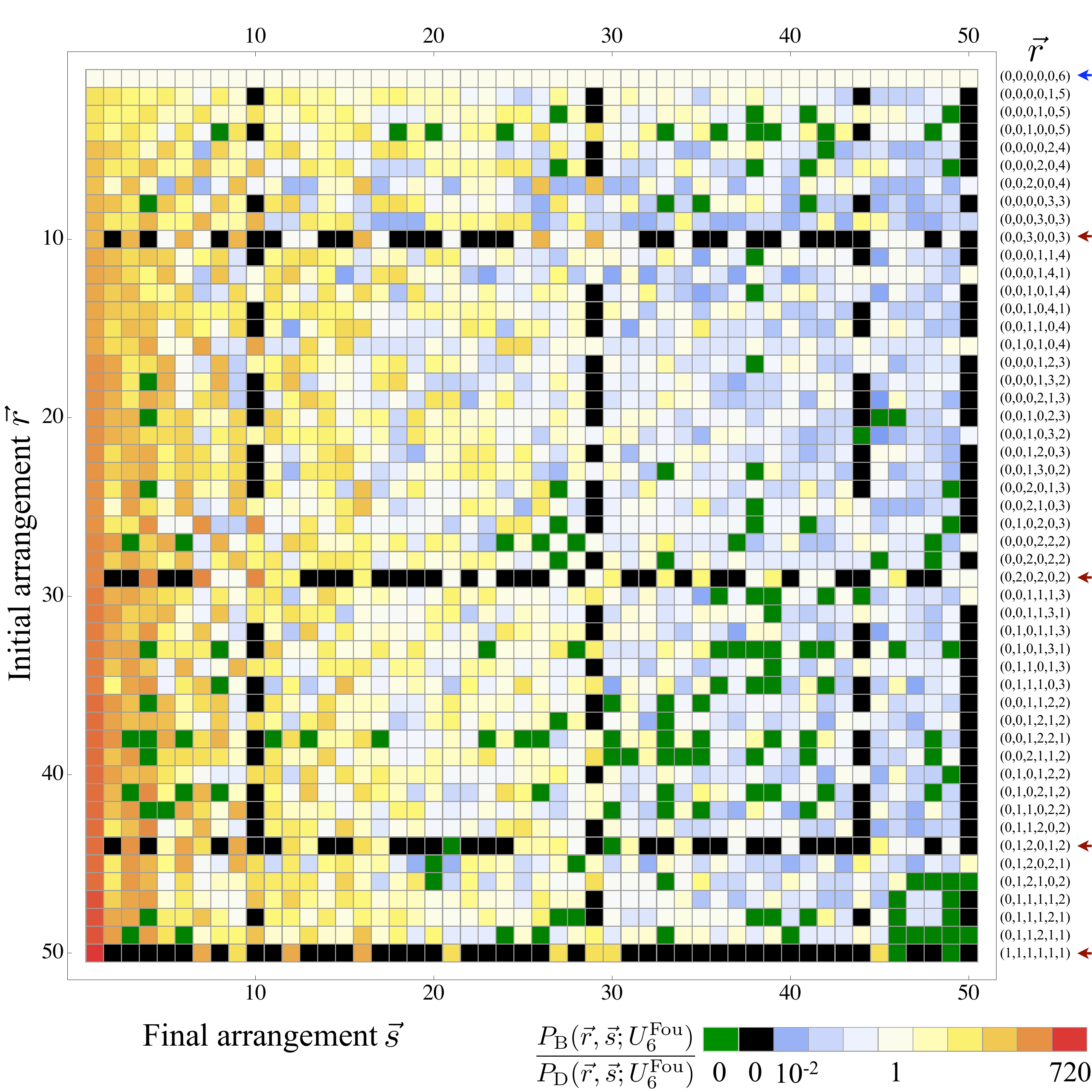} 
\caption{ {\bf Many-boson interference in a setup with $N=6$ bosons and $n=6$ modes.} The color coding indicates the quotient of the probability for bosons, $P_{\text{B}}(\vec r, \vec s; U^{\text{Fou}})$, Eq.~(\ref{BoseFourier}), to the classical probability $P_{\text{D}}(\vec r, \vec s; U^{\text{Had}})$, Eq.~(\ref{distsimple}), as a function of the input state $\vec r$ and of the output state $\vec s$. The input- and output configurations are arranged in the same order. Black fields denote transitions that are suppressed due to the suppression law (\ref{law}), while green fields represent suppressed events which are not predicted by the law.  One can identify the periodic initial and final states (marked with red arrows) as the black horizontal and vertical structures. For the initial states $\vec r=(0,0,3,0,0,3)$ and $\vec r=(0,1,2,0,1,2)$, the \emph{same} final arrangements $\vec s$ are suppressed, since the suppression law only depends on the period length of the initial state. As an exception, the transition $\vec r=(0,1,2,0,1,2)$ to $\vec s=(0,2,0,2,0,2)$ is suppressed, whereas for $\vec r=(0,0,3,0,0,3)$, it is not. This is rooted in the period of  $\vec s=(0,2,0,2,0,2)$, $m=2$, and $2\cdot \sum_j d_j((0,1,2,0,1,2))=50 \neq 0 \text{ mod } 6$, {\it i.e.}~the transition is suppressed owing to the \emph{reverse} of the law, which is obtained by exchanging the input and output states. The behavior of $\vec r=(0,0,3,0,0,3)$ and $\vec r=(0,1,2,0,1,2)$ for the remaining, unsuppressed transitions differs strongly.  The first line (marked by a blue arrow) corresponds to the initial state $\vec r=(0,0,0,0,0,6)$, which does not exhibit any interference effects, since all particles are initially in the same single-particle state and no competing many-particle paths with different phase arise. A hierarchy of initial states can be observed for the final state $\vec s=(0,0,0,0,0,6)$ (outmost left column): The more spread out the particles are in the input modes, the stronger is the enhancement of the final state due to bosonic bunching. } \label{appl6} \end{figure*}

With a scenario of $N=6$ bosons in an $n=6$ mode setup, we exemplify the impact of many-particle interference and of the suppression law (\ref{law}). An overview of the system is given in Fig.~\ref{appl6}, where the quantum enhancement, {\it i.e.~}the bosonic probability $P_\text{B}(\vec r, \vec s; U_6^{\text{Fou}})$ (Eq.~(\ref{BoseFourier})) divided by the classical probability $P_{\text{D}}(\vec r, \vec s; U_6^{\text{Fou}})$ (Eq.~(\ref{distsimple})), is displayed. Constructive many-particle interference thus leads to a quantum enhancement  larger than unity (reddish colors), whereas a value smaller than unity (blueish colors) indicates destructive interference. Fully suppressed events are marked in black (green) when they are (not) predicted by the suppression law (\ref{law}).

The arrangements are ordered according to their occupancy, such that final (initial) arrangements with many particles in few modes are found on the left (upper) part of the plot. Arrangements which can be related to each other via cyclic permutation or inverse are omitted (they lead to identical transition probabilities), such that 50 arrangements remain. Initial states with a period (marked with a red arrow) lead to many fully suppressed transitions due to the suppression law, whereas the remaining unsuppressed events are, consequently, typically enhanced.

A coarse-grained trend emerges: Events with few, highly occupied modes (on the left side) tend to be enhanced, events for which the particles are well distributed over the modes (on the right side) are rather suppressed. A closer look reveals that this \emph{average} trend has many exceptions: Suppressed transitions can be found within the events with many particles in few modes, and enhanced transitions with many occupied modes appear; fully destructive interference occurs within the highly and within the sparsely occupied arrangements. In general, the behavior of an event is sensitive to the exact state preparation and on the final state configuration: Shuffling one particle into a neighboring mode often turns constructive interference into destructive, and vice-versa. Similarly, the interference pattern shown in Fig.~\ref{appl6} strongly depends on the phases of the matrix (\ref{foumadef}). Phase variations immediately impact on all transition amplitudes besides singular cases in which only one many-particle path is possible, such as for the initial or final state $(6,0,0,0,0,0)$.

\subsubsection*{Mixed initial states and the emergence of bosonic behavior}
The interference effects in Fig.~\ref{appl6} are sensitive to the loss of indistinguishability of the particles. When the latter is not ensured anymore, their interference capability is jeopardized, such that the resulting transition probabilities eventually approach the ones for distinguishable particles \cite{fourphotons}. 

The interference pattern also depends on the coherence of the many-particle wavefunction, {\it i.e.}~on the purity of the initial state and on the coherence of the time evolution. In order to see this, we consider a state in which the individual particles are still fully indistinguishable, while they are prepared in a mixture of all possible nonequivalent arrangements, ({\it i.e.}~arrangements that cannot be related to each other via cyclic symmetry), 
\eq 
\rho_{\text{mix}}= \frac{1}{N_{\text{arr}}^{\text{B}}} \sum_{\vec r} \ket{\Psi(\vec r)}\bra{\Psi(\vec r)} , \label{mixedstate}
\en
where $N^{\text{B}}_{\text{arr}}$ is the total number of arrangements given by Eq.~(\ref{numarrbos}). The probability for a final arrangement $\vec s$ for $\rho_{\text{mix}}$ amounts to 
\eq 
P(\rho_{\text{mix}},\vec s) &=& \frac{1}{N_{\text{arr}}^{\text{B}}} \sum_{\vec r} P(\vec r, \vec s; U^{\text{Fou}}_6) \nonumber \\ 
 &=& \langle P(\vec r, \vec s; U^{\text{Fou}}_6) \rangle_{\vec r} , \label{probmix}
\en
{\it i.e.}~to the transition probability averaged over the nonequivalent initial states $\vec r$. The interference effects that govern the behavior for each pure initial state $\vec r$ are averaged out for $\vec \rho_{\text{mix}}$, and all final arrangements become approximately equally probable; we expect 
\eq \langle P(\vec r, \vec s; U^{\text{Fou}}_6) \rangle_{\vec r} \approx  P_{\text{E}}(\vec s, U_6^{\text{Fou}}):=\frac 1 {N^{\text{B}}_{\text{arr}}} . \label{defesti} \en

\begin{figure} \center
\includegraphics[width=\linewidth,angle=0]{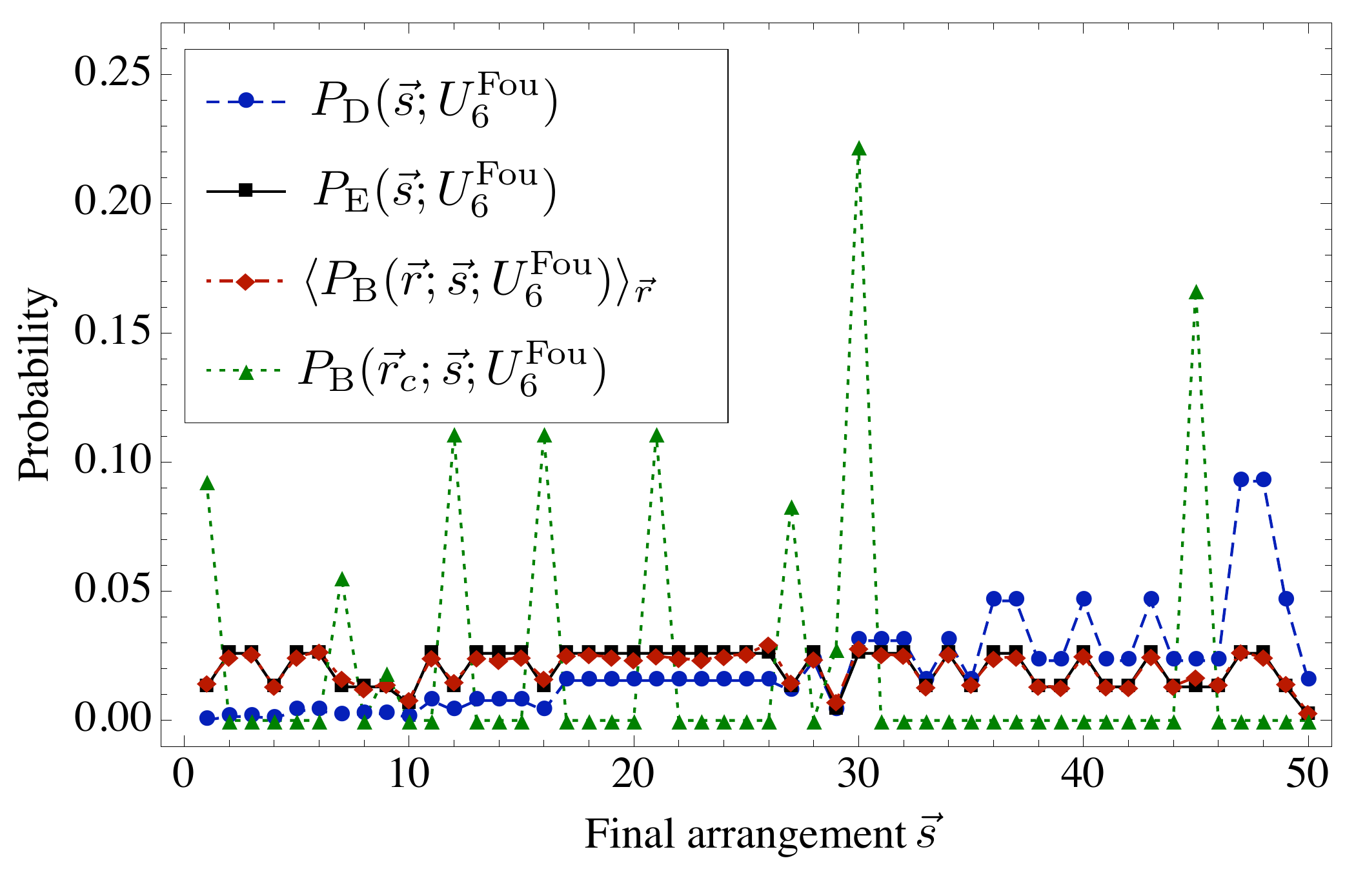} 
\caption{ {\bf Behavior of pure and mixed states of $N=6$ bosons that are scattered in an $n=6$ mode setup.} We show the probability for final arrangements $\vec s$, arranged in the same order as in Fig.~\ref{appl6}, for distinguishable particles, $P_{\text{D}}$ (blue  circles, dotted line, Eq.~(\ref{distsimple})), for the equiprobable distribution $P_E=1/N_{\text{arr}}^{\text{B}}$ (black  squares, solid line, Eq.~(\ref{defesti})), for the mixed state of bosons $\rho_\text{mix}$ (Eq.~(\ref{mixedstate})) to which all possible initial arrangements contribute with the same weight, {\it i.e.}~$\langle P_\text{B}(\vec r, \vec s; U^{\text{Fou}}_n) \rangle_{\vec r}$ (red diamonds, dash-dotted line, Eq.~(\ref{probmix})), and for bosons that are prepared in the initial state $\vec r_c=(1,1,1,1,1,1)$, $P_\text{B}(\vec r_c, \vec s; U^{\text{Fou}}_6)$ (green triangles, dotted line). Since we group each arrangement $\vec s$ together with its cyclically permuted and mirrored counterparts, $P_E(\vec s)$ is not a constant, but it reflects the multiplicity of equivalent arrangements. The trend to favor arrangements with large populations is visible by comparing the probability for distinguishable particles $P_{\text{D}}$ to the estimate $P_{\text{E}}$, which favors multiply occupied states. The exact calculation for bosons in $\vec r_c$, however, does not exhibit this trend at all, and the picture is dominated by many-particle interferences. Only when the mean over the initial states is performed, as in Eq.~(\ref{mixedstate}), we lose many-particle interference and recover  a clear bosonic bunching tendency, reflected by the values of $\rho_\text{mix}$. }\label{mixfigure} \end{figure}

We verify this statistical argument by comparing the event probability for distinguishable particles, $P_\text{D}(\vec s, U^{\text{Fou}})$, the probability for bosons prepared in the arrangement $\vec r_c=(1,1,1,1,1,1)$, the estimated probability $P_{\text{E}}$, and the probability given the mixed initial state $\rho_{\text{mix}}$ in Fig.~\ref{mixfigure}. When the particles are prepared in $\vec r_c$, {\it i.e.}~in a pure state, the system exhibits interference, and the resulting event probabilities strongly differ from the combinatorial value (\ref{distsimple}) for distinguishable particles. In comparison to the latter, events with many particles per mode (left side) are enhanced, events with many occupied modes (right side) are rather suppressed in the equiprobable distribution $P_{\text{E}}$ given in Eq.~(\ref{defesti}). Finally, the probability distribution for the state  $\rho_{\text{mix}}$,  Eq.~(\ref{probmix}), matches $P_\text{E}$ very well, which confirms the approximation (\ref{defesti}). Due to the mixed preparation of the particles, no interference pattern can crystallize for $\rho_{\text{mix}}$, and only an average, bosonic behavior remains. 

In other words, we recover the well-known statistical behavior of bosons when the \emph{purity} of the initial state is destroyed, while the behavior of distinguishable particles emerges when the \emph{indistinguishability} of the particles is lost. 

\subsection*{Boson-fermion comparison}
When we restrict ourselves to initial and final states with at most one particle per mode, which we name \emph{Pauli states} in the following, we can compare distinguishable particles, bosons, and fermions. 

In any setup with two particles, exactly two two-particle paths contribute to the total transition amplitude between any initial and any final Pauli state, which is reflected by the two terms in the sum (\ref{generalampli}). The two amplitudes are summed for bosons and subtracted for fermions, which leads to an antipodal behavior. For three or more particles, however, more paths contribute to the amplitude in (\ref{generalampli}), and no dichotomy is observed.

Indeed, when we focus on the Fourier matrix (\ref{foumadef}), the \emph{very same} transitions are predicted to be suppressed for an odd number of bosons and fermions, according to Eq.~(\ref{law}) and (\ref{lawfermi1}). This emphasizes that our criterion for destructive interference does not rely on the (anti)commutativity of (fermionic) bosonic operators, but rather on the coherent superposition of many-particle paths. 

At first sight, it seems that the suppression law for even particle numbers and odd $N/p$ (\ref{lawfermi2}) predicts an anti-podal behavior of bosons and fermions. For exactly two particles, any arrangement with cyclic symmetry has indeed $m=n/2, p=2$, and transitions that are suppressed for bosons are enhanced for fermions, and vice-versa. In general, transitions with \eq Q:=\text{mod}\left( m \sum_{j=1}^n d_j(\vec s),n \right) \in \{ 0, n/2 \} , \en are only suppressed for one species, and not necessarily for the other.  For large $n$, however, most transitions naturally lead to values of $Q$ different from 0 and $n/2$. Consequently, fermions and bosons also share many suppressed events for even particle numbers, unless $N=2$. 

\begin{figure*} \center
\includegraphics[width=\linewidth,angle=0]{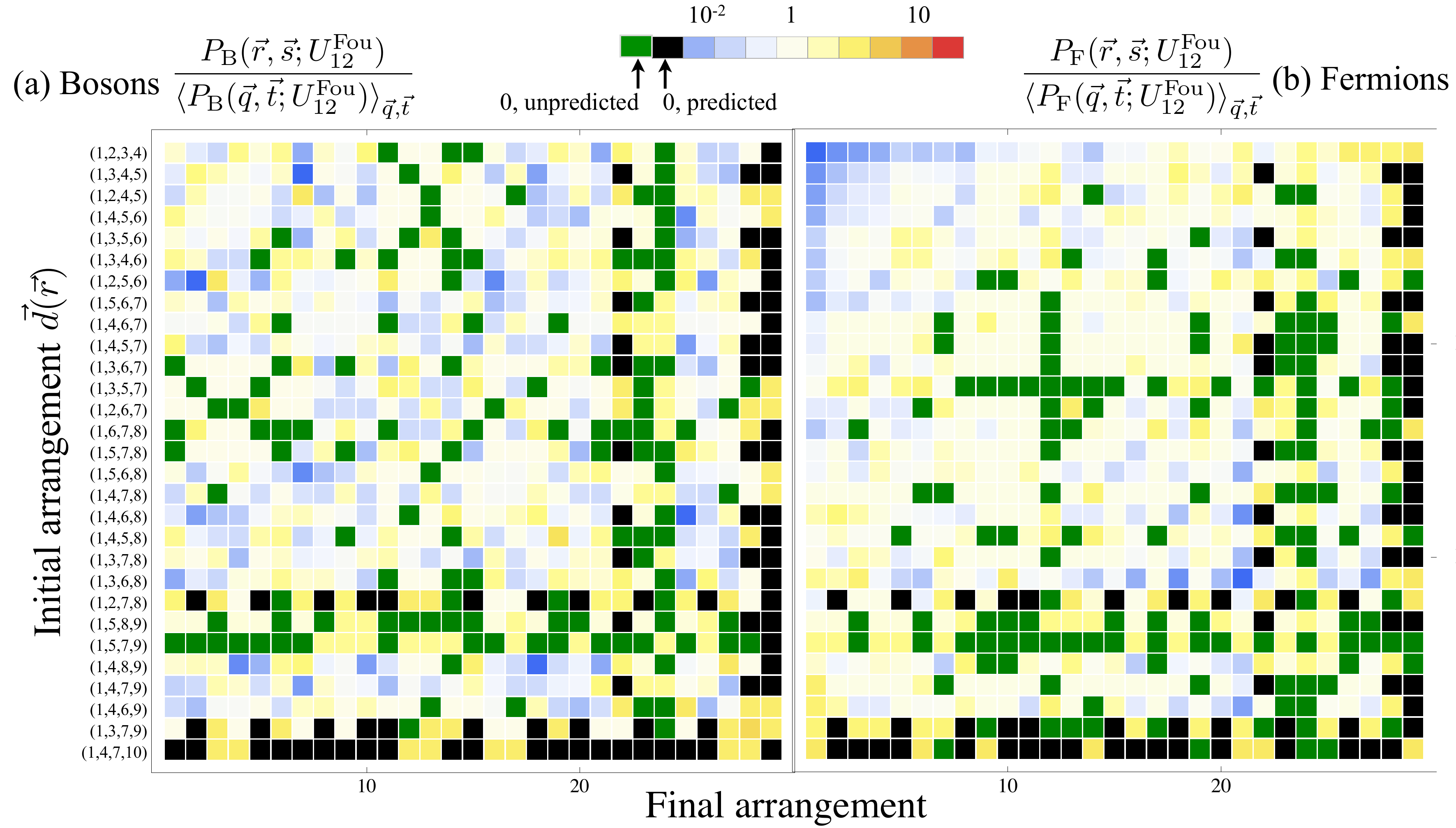} 
\caption{{\bf Many-boson (a) and many-fermion (b) interference of $N=4$ particles in a $n=12$-mode setup. } The color code reflects the quotient of the event probability for bosons (a) and fermions (b), divided by the average probability to find a Pauli state. The arrangements are given by their mode assignment list, the order of the final arrangements is the same as for the initial arrangements. Three out of the 29 inequivalent states possess a cyclic symmetry. The arrangements with elementary mode occupation lists $\vec k=(0,0,0,0,1,1)$ and $\vec k=(0,0,0,1,0,1)$ have period length $m=6$, such that the same transitions are predicted to be suppressed for bosons and fermions; Eqs.~(\ref{law}) and (\ref{lawfermi1}) apply. In contrast, the arrangement based on $\vec k=(0,0,1)$ has period length $m=3$. Since $\nicefrac{N}{p}=1$, (\ref{lawfermi2}) applies.   }\label{BoseFermiCompare} \end{figure*}

\begin{figure} \center
\includegraphics[width=\linewidth,angle=0]{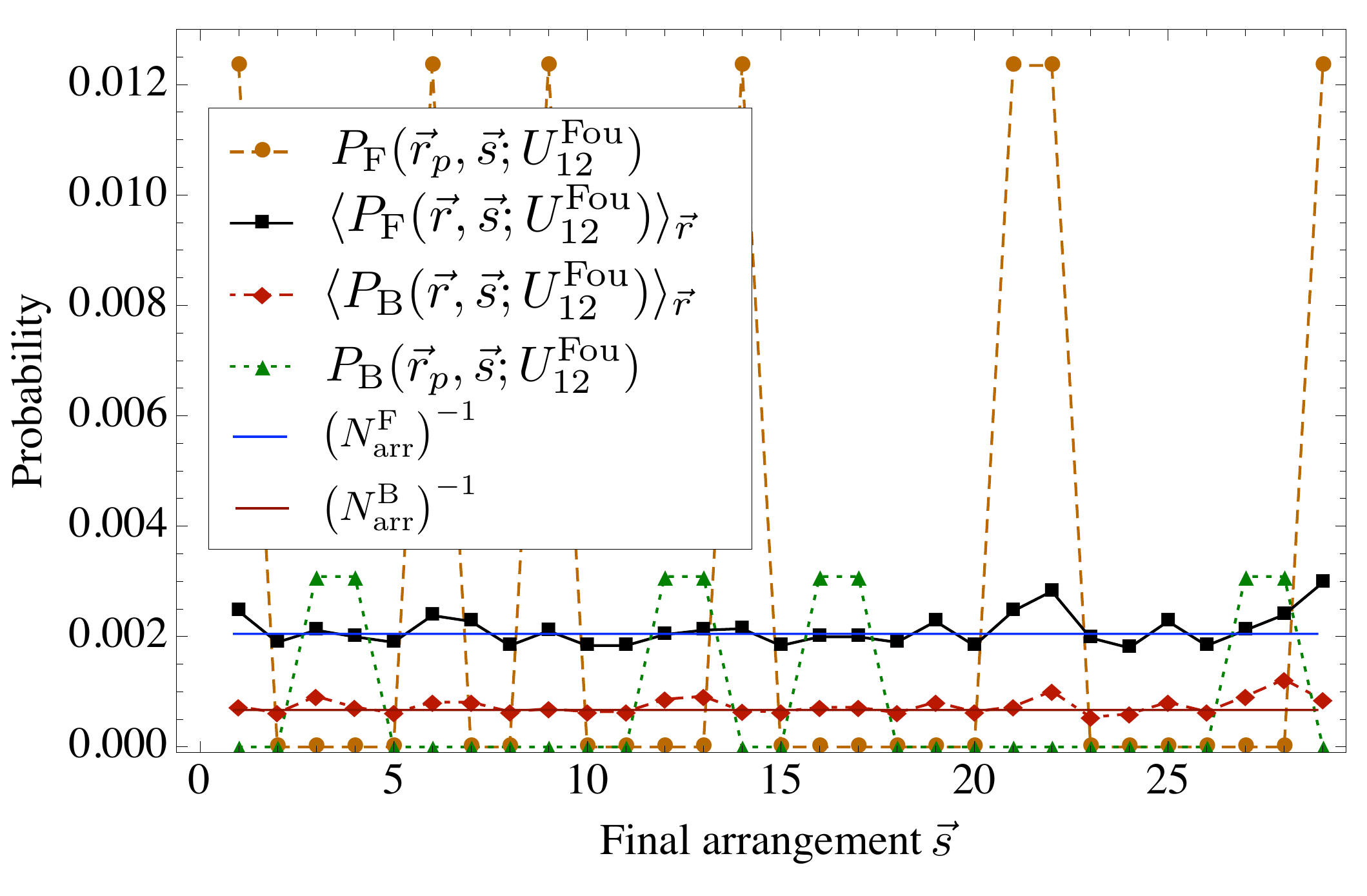} 
\caption{{\bf Many-boson and many-fermion scattering of pure and mixed states.} The probabilities for final arrangements $\vec s$ are arranged in the same order as in Fig.~\ref{BoseFermiCompare}. Bosons (green triangles, dotted line) are compared to Fermions (brown circles, dashed line) for the initial state $\vec d(\vec r_p)=(1,4,7,10)$ that corresponds to the very last line in Fig.~\ref{BoseFermiCompare}. An average is taken over all nonequivalent initial states, which results in the black squares for fermions and red diamonds for bosons. It approaches the inverse of the total number of states for fermions (blue solid line) and bosons (red solid line), respectively.  }\label{FermiCompareAveraged} \end{figure}

\subsubsection*{Mean bosonic and fermionic behavior}
Despite the similarities between fermionic and bosonic interference, a \emph{mean} trend for the probability to yield a Pauli state appears: Since events other than Pauli states  are strictly forbidden for fermions, the average probability for Pauli states is enhanced with respect to distinguishable particles. For bosons, events with higher occupation are favored, and Pauli states are rather suppressed. 

In a Bell multiport with $n=12$ modes and $N=4$ particles, we find this tendency: The probability for a Pauli state amounts to 1/864 for distinguishable particles, according to Eq.~(\ref{distsimple}).  The average fermionic probability is enhanced, it amounts to the inverse of the number of available states, Eq.~(\ref{numarrbosf}),  $\left(N_{\text{arr}}^{\text{F}}\right)^{-1}=1/495$. For bosons, events with multiple occupation are privileged, the average probability for a Pauli state becomes $\approx 7.50\cdot10^{-4}$, 
 which is close to the inverse of the number of accessible states, $\left( N_{\text{arr}}^{\text{B}} \right)^{-1}=1/1365\approx 7.33\cdot 10^{-4}$.

From this general trend, however, no systematic statement can be inferred for the individual transitions, which are displayed in Fig.~\ref{BoseFermiCompare}: No anti-correlation between fermionic and bosonic behavior is visible, and the  correlation coefficient between the enhancement or suppression of fermionic/bosonic events amounts to only -0.05. The enhancement of a fermionic transition does not imply the suppression of the corresponding bosonic transition.

Decreasing the particle density, {\it i.e.}~increasing the number of modes $n$ for a constant number of particles $N$, has no impact on the strength of  interference effects: Fully suppressed  and highly enhanced transitions remain, and so does the general impact of many-particle interference.  In turn, the difference between the \emph{average} bosonic/ferminionic behavior fades away when the number of modes is increased: The quotient of bosonic to fermionic accessible states, $N_{\text{arr}}^{\text{B}}/N_{\text{arr}}^{\text{F}}$, given by Eqs.~(\ref{numarrbos}) and (\ref{numarrbosf}), approaches unity for $n\rightarrow \infty$, and the average statistics then resembles the combinatorial distribution of distinguishable particles. This reflects the absence of the Fermi pressure \cite{Romer:1994uq} and of bosonic behavior for low-density systems.

\subsubsection*{Mixed initial states}

Just like for the case of six bosons discussed above, the transition amplitudes for fermions and bosons that are prepared in Pauli states depend on the initial pure preparation of the particles: Given a fully mixed state analogous to (\ref{mixedstate}), but restricted to Pauli states, the resulting event probability approaches a constant value that does not depend strongly on the final arrangement. Since the indistinguishability of the particles is not jeopardized by the mixedness of the initial state, the Pauli principle prevails for fermions, and so does the average privilege of multiply occupied states for bosons. The probability for any final state then approaches the inverse of the number of states, $\left(N_{\text{arr}}^{\text{F}/\text{B}} \right)^{-1}$, which were given in Eqs.~(\ref{numarrbos}, \ref{numarrbosf}). The behavior of bosons and fermions is compared in Fig.~\ref{FermiCompareAveraged}, where the loss of interference due to the mixedness of the initial state is apparent. 

In analogy to the bosonic case, the loss of the purity of an initial state of fermions leads to the loss of the interference pattern, while the quantum statistical behavior persists. Only when the indistinguishability of fermions is jeopardized, also the impact of the Pauli principle  fades away. 

\FloatBarrier

\section{Conclusions and Outlook}
The coherent many-particle evolution of non-interacting identical particles is governed by many-particle interference, with consequences that go far beyond bosonic (fermionic) (anti)bunching encountered in incoherent environments. The latter phenomena can be understood from the postulate that assigns every micro-state the same realization probability \cite{Romer:1994uq} when the specific constraints to bosons and fermions are respected. The (anti)symmetrization, however, establishes also a many-particle \emph{coherence} property that leads to the encountered interference effects, which seldom reproduce the familiar statistical behavior.  

Although any many-particle interference setup can be treated by Eq.~(\ref{generalampli}), the analysis becomes rather tedious when the number of particles and modes are not both small, which is due to the computational complexity inherent in Eqs.~(\ref{bosontranspermurep}) and (\ref{determinantferm}), and also due to the large number of accessible states, Eqs.~(\ref{numarrbos}), (\ref{numarrbosf}). For Bell multiport beam splitters, we can circumvent these difficulties by exploiting the available symmetries, which allows the systematic confrontation of many-boson to many-fermion scattering. Since for any number of particles $N$ and any number of modes $n$ with a non-trivial greatest common divisor, {\it i.e.}~$GCD(N,n)>1$, non-trivial periodic states of the form (\ref{defiperiodicstate}) can be found, numerous applications of the  suppression laws are possible. We thus have a constructive recipe for many-particle large-visibility setups at our hands, since events that exhibit destructive interference can be found for arbitrary system sizes, whereas the explicit evaluation of (\ref{generalampli}) becomes prohibitive already for moderate particle and mode numbers. This  provides a powerful characterization toolbox for the indistinguishability of many particles, for the purity of their initial state preparation and for the many-particle coherence of the time-evolution -- which are all imperative properties for quantum technologies. 

The interference patterns shown in Figs. \ref{appl6}-\ref{FermiCompareAveraged} are specific to the Bell multiport, but also exemplary for any other scattering scenario. Our results suggest a forceful distinction between \emph{quantum statistical effects}, {\it i.e.}~the (anti-)bunching of  uncontrolled (fermions) bosons in a thermal state \cite{Jeltes:2007ly} or in the mixed state (\ref{mixedstate}), and \emph{many-particle interference}, {\it i.e.}~the coherent conspiration of many-particle paths. The former are the consequence of the interplay of the kinematic constraints with the statistical uncertainty, whereas the latter are jeopardized by such uncertainty. A many-particle path picture is also applicable in the incoherent case \cite{Jeltes:2007ly,Folling:2005fk}. However, one then only distills an average, incoherent statistical behavior that can also be understood from the equal a priori probability postulate \cite{Romer:1994uq}.  

To name a further difference, quantum statistical effects fade away with decreasing particle density, in contrast to many-particle interference. So far, the literature has mainly concentrated on the evolution of two-particle states \cite{Hong:1987mz} and on two-point correlation functions \cite{Henny:1999cr,Jeltes:2007ly}, for which these phenomena appear to be synonymous: The Hong-Ou-Mandel effect \cite{Hong:1987mz} is in ostensible agreement with the bunching behavior of thermal bosons \cite{Hanbury-Brown:1956vn}, and it also reflects a strong boson-fermion dichotomy \cite{PhysRevA.58.4904}. As we have shown, this ostensibly intuitive picture needs to be abandoned immediately as soon as more than two particles are considered. 

The loss of purity in the initial state leads to a deterioration of interference effects and to a recovery of an average bosonic or fermionic behavior, while a loss of mutual particle indistinguishability leads to the combinatorial behavior of distinguishable particles. In general, decoherence is thus a more diverse and complex phenomenon  in the many-particle domain than in the single-particle realm. It remains to be studied how decoherence processes affect many-particle interference: We speculate that different mechanisms deteriorate the interference pattern in different ways, such that one recovers -- in the limit of strong decoherence -- either the behavior of fully distinguishable particles \cite{fourphotons,Ra:2011ys} or a bosonic/fermionic statistical behavior.  \vspace{1cm}

{\it Acknowledgements} The authors are grateful to Klaus M\o lmer for enlightening discussions. M.C.T. acknowledges support by German National Academic Foundation. A.B. acknowledges partial support through the EU-COST Action MP1006 "Fundamental Problems in Quantum Physics".

\appendix

\section{Proof of the suppression laws}

For bosons and fermions, by inserting the definition of the Fourier matrix (\ref{foumadef}), we can rewrite the transition amplitudes (\ref{bosontranspermurep}) and (\ref{determinantferm}) as sums of amplitudes with unit modulus: 
\begin{widetext}
\eq 
P_{\text{B/F}}(\vec r, \vec s; U^{\text{Fou}})&=& \frac{1}{\prod_{k=1}^n r_k! s_k!} \frac {1} {n^N} \left| \sum_{\sigma \in S_N } \text{sgn}_{\text{B/F}}(\sigma) \text{exp}\left( i \frac {2\pi}{n} \sum_{j=1}^N d_{\sigma(j)}(\vec r) \cdot d_j(\vec s) \right) \right|^2 , \label{BoseFourier} 
\en
where $ \text{sgn}_{\text{B}}(\sigma)=1$ and $ \text{sgn}_{\text{F}}(\sigma)= \text{sgn}(\sigma)$.
\end{widetext}

\subsubsection*{Bosons}
With the definition 
 \eq \Theta(\vec r, \vec s, \sigma)=\sum_{j=1}^N d_{\sigma(j)}(\vec r) d_j(\vec s) \label{thetadef}, \en 
  the probability $P_{\text{B}}(\vec r, \vec s; U^{\text{Fou}})$ in (\ref{BoseFourier}) becomes proportional to 
\eq \left| \sum_{\sigma \in S_N } \text{exp}\left( i \frac {2\pi}{n}\Theta(\vec r, \vec s, \sigma ) \right) \right|^2 . \label{BoseFourierTheta}\en
In other words, the natural number $\Theta(\vec r, \vec s, \sigma)$ is the total phase that is acquired by the many-particle wavefunction (in multiples of $\nicefrac{2 \pi}{n}$), when one specific many-particle path defined by $\sigma$, from the initial state $\vec r$ to the final state $\vec s$,  is realized.  Since $\Theta(\vec r, \vec s, \sigma)$ is a natural number, the sum in (\ref{BoseFourierTheta}) contains only $n$-th roots of unity, 
\eq P_{\text{B}}(\vec r, \vec s; U^{\text{Fou}})&\propto & \left| \sum_{k=0}^{n-1} c_{k} e^{i \frac{2 \pi}{n} k } \right|^2,  \label{rootsofunity} \en  with  $ c_k \in \mathbbm{N}, ~ \sum_{k=0}^{n-1} c_k =N! $,  and where the value of $c_k$ corresponds to the number of permutations $\sigma$ for which  $\mod(\Theta(\vec r, \vec s, \sigma),n)=k$ \cite{Graham:1976nx}: 
\eq u_{k}(\vec r, \vec s)&=& \left\{ \sigma | \Theta(\vec r,\vec s, \sigma) = k \text{ mod } n \right\} , \nonumber \\
c_k &=& |u_k| , \label{definitionpermutsizes}
\en
where $x=y \text{ mod } k$ means that there is a integer number $l$ such that $x=l\cdot k + y$. Hence, the sets $u_k$ group all many-particle paths that acquire the same phase, $\nicefrac{2 \pi k}{ n}$. Since all moduli of involved probabilities are equal, these paths possess the same \emph{amplitude}. It is helpful to define \eq Q(m, \vec s) =\text{mod}\left(m \sum_{l=1}^N  d_l(\vec s) , n\right) ,\en
and an operation $\Gamma$ on the permutations $\sigma$, 
\eq \Gamma(\sigma)(k) = \left( \sigma(k) + \frac N p \right)\mod n ,  \label{shifter}\en
which shifts the permutation $\sigma$ by the number of particles in each period repetition, $N/p$. Due to the $m$-periodicity of the initial state $\vec r$, which implied (\ref{propper}), the value of the total acquired phase $\Theta(\vec r, \vec s, \sigma)$ acquires the constant $Q(m, \vec s)$ when the above transformation is applied on a permutation $\sigma$:
\eq  \Theta\left(\vec r, \vec s, \Gamma(\sigma) \right) &=& \sum_{j=1}^N d_j(\vec s) d_{\Gamma(\sigma(j))}(\vec r) .\en 
Using (\ref{propper}) and (\ref{shifter}), we find 
\eq  \Theta\left(\vec r, \vec s, \Gamma(\sigma) \right) & =& \sum_{j=1}^N d_j(\vec s) d_{\sigma(j)}(\vec r) + m \sum_{j=1}^N d_{j} (\vec s) \nonumber \\&=&  \Theta\left(\vec r, \vec s, \sigma \right) + Q(m, \vec s)  \text{ mod } n . \en
Thus, if $Q(m, \vec s) \neq 0$, the repeated application of $\Gamma$ gives us a bijection between all pairs of $u_{b+a \cdot  Q(m, \vec s)}$ for $b\in \left\{1,..,n \right\} $ and $a \in \left\{0,1,..,n-1 \right\}$. 
Therefore, we have
 \eq 
 \forall b  \in \{0,\dots n-1\}, \forall a \in \mathbbm{N} :  c_{b+a Q(m, \vec s)} =c_{b} ,   \label{symmprobbos} \en
Since only index values mod $n$ are relevant, all equalities that can be inferred from (\ref{symmprobbos}) are contained by the following $n$ equalities:
 \eq 
 \forall b  &\in& \left\{0,\dots \frac{n \cdot Q(m, \vec s)}{\text{LCM}\left(Q(m, \vec s),n \right)}-1 \right\},  \label{symmprobbos2} \\  \forall a &\in & \left\{0, \dots \frac{\text{LCM}\left(Q(m, \vec s), n\right)}{Q(m, \vec s)} -1 \right\} :  c_{b+a Q(m, \vec s)} =c_{b} ,  \nonumber \en
where $\text{LCM}(x,y)$ denotes the \emph{least common multiple} of $x$ and $y$. Since \eq \text{LCM}(x,y) \text{GCD}(x,y)= x \cdot y, \en $b$ can effectively take $GCD(Q(m,\vec s), n)=:g$ distinct values, and each $c_b$ is equal to $\nicefrac n g-1$ other coefficients $c_{b+a Q}$. By setting the summation index $k$ in (\ref{rootsofunity}) to $k=b+a \cdot Q$, the sum (\ref{rootsofunity}) can be rewritten as 
\eq P_{\text{B}}(\vec r, \vec s; U^{\text{Fou}})\propto \left| \left( \sum_{b=0}^{g-1} c_b e^{i \frac{2 \pi}{n} b } \right) \left(  \sum_{a=0}^{\nicefrac n g-1} e^{i \frac{2 \pi}{n} Q(m, \vec s) \cdot a}  \right) \right|^2 , \label{circlescomplex} \en
where we exploited explicitly (\ref{symmprobbos}), such that the sum factorizes in two parts. 
The sum over $a$ is a truncated geometric series, 
\eq \sum_{j=0}^{l-1} x^j = \frac{1-x^l}{1-x} ,  \label{geomser} \en
 with $l=\nicefrac n g$ and $x=e^{i \frac{2\pi}{n} Q(m,\vec s)}$. Since $x$ is a $n$-th root of unity and $x\neq 1$ (since we assumed $Q(m ,\vec s) \neq 0$), (\ref{geomser}) vanishes, and so does (\ref{circlescomplex}).

\subsubsection*{Adaptation for fermions} \label{adaptfermions}
In order to adapt the suppression law for fermions, we need to include the signature of the respective permutation in the sum of the amplitudes (\ref{BoseFourier}). In analogy to (\ref{rootsofunity}), we find 
\eq P_{\text{F}}\left(\vec r, \vec s; U^{\text{Fou}} \right) \propto \left| \sum_{\sigma \in S^E_N} e^{i \frac{2 \pi}{n} \Theta(\vec r, \vec s, \sigma ) }  -  \sum_{\sigma \in S^O_N} e^{i \frac{2 \pi}{n}  \Theta(\vec r, \vec s, \sigma ) } \right|^2 \label{newsumfermi}, \en
where we split the sum (\ref{BoseFourier}) into even permutations $S_N^E$, and odd ones $S_M^O$. We now need to consider the sets of even (E) and odd (O) permutations with $\Theta(\vec r, \vec s, \sigma)=k$, 
\eq 
u_b^{E(O)} (\vec r, \vec s) = \hspace{4.5cm} \label{fermionspermuss}  \\
\{ \sigma | \Theta(\vec r, \vec s, \sigma) = b \text{ mod } n, \text{sgn}(\sigma)=+1 (-1) \} ,  \nonumber \en 
separately. Their respective cardinality is denoted by $c_b^{E(O)}=|u_b^{E(O)}|$. We also need to infer the action of $\Gamma$ on the parity of the permutations. 

\subsubsection{Odd particle number $N$, or even $\nicefrac{N}{p}$}
For odd particle numbers $N$, the application of $\Gamma$ on a permutation $\sigma$ does not change its parity: Any cyclic permutation on a set of odd size (here the set of indices $\{1, \dots , N\}$, of size $N$) is itself an even permutation because it is composed of the even number $N-1$ of (odd) elementary transpositions. 

The parity of a permutation remains also unchanged under the application of $\Gamma$ when the number of particles $N$ is even and $\nicefrac{N}{p}$ is also even. The application of $\Gamma$ on a permutation corresponds to an $\nicefrac{N}{p}$-fold cyclic shift that is composed by $\nicefrac{N}{p}$ (odd) elementary shifts. Its parity is therefore even, since $N/p$ is even. 

In these two cases, we find, as for bosons but now \emph{independently} for even and odd permutations:
 \eq  \forall b \in \{0,\dots n-1\}:  c^E_{b+Q} =c^E_{b} , \  c^O_{b+Q} =c^O_{b} .\en
 The symmetry property of (\ref{symmprobbos}) is hence inherited \emph{independently} by the even and the odd permutations. The formulation and the consequence of the suppression law consequently remain unchanged: Formally, it states that the even and odd part of the sum (\ref{newsumfermi}) \emph{both} vanish. 

\subsubsection{Even particle numbers $N$ and odd $\nicefrac{N}{p}$}
For even $N$ and odd $\nicefrac{N}{p}$, 
 the application of $\Gamma$ onto a permutation $\sigma$ changes its signature, and we find 
 \eq  \forall b \in \{0,\dots n-1\}:  c^E_{b+Q} =c^O_{b} ,\ \   c^O_{b+Q} =c^E_{b} . \label{varilaw} \en
The signatures of the permutations are thus interchanged, and our suppression law in the formulation (\ref{law}) is not valid anymore. By a case-by-case analysis, we can explore the consequences of (\ref{varilaw}):
\begin{itemize}
\item The condition $\text{mod}(Q,n)=0$ is sufficient for the full suppression of the respective transition. In this case, $ \forall b: c^E_{b}=c^O_{b} $ and each amplitude $c_b^E e^{i \frac{2\pi}{n} b}$ has an amplitude $-c_b^O e^{i \frac{2\pi}{n} b}$ equal in magnitude, but opposite in sign. A transition between $\vec r$ and $\vec s$ that is not necessarily suppressed for bosons becomes so for fermions.
\item A transition with $\text{mod}(Q,n)= \nicefrac{n}{2}$ is not necessarily suppressed, since $c^E_{b}=c^O_{b+\frac n 2}$, such that two amplitudes $c^E_{b}$ and $c^O_{b+\frac n 2}$ do not cancel -- as they do for bosons where they lie on opposite sides of the origin in the complex plane -- but instead they \emph{enhance} each other. 
\item For all other values of $\text{mod}(Q,n)$, {\it i.e.~}values that are neither zero nor $\nicefrac{n}{2}$, the respective transitions are, again, necessarily suppressed, since (\ref{circlescomplex}) turns into 
\eq 
P_{\text{F}}(\vec r, \vec s; U^{\text{Fou}})\propto \left|  \left( \sum_{r=0}^{g-1} \left( c_b^{E} -c_b^{O} \right) e^{i \frac{2 \pi}{n} b } \right) \times \right. \nonumber \\ 
\left. \left( 
\sum_{a=0}^{\nicefrac{n}{g}-1} (-1)^a e^{i \frac{2 \pi}{n} Q(m, \vec s) \cdot a} 
\right) \right|^2 , \en
where the sum over $a$ can, again, be represented as truncated geometric series and vanishes since $n$ is even by assumption.  \vspace{.25cm}
\end{itemize}


\begin{thebibliography}{10}

\bibitem{Girardeau:1965ys}
M.~D. Girardeau,
\newblock Phys. Rev. {\bf 139}, B500 (1965).

\bibitem{Romer:1994uq}
H.~R{\"o}mer and T.~Filk,
\newblock {\em Statistische Mechanik} (Wiley-VCH, Weinheim, 1994).

\bibitem{Hanbury-Brown:1956vn}
R.~Hanbury~Brown and R.~Q. Twiss,
\newblock Nature {\bf 177}, 27 (1956).

\bibitem{Hodgman25022011}
S.~S. Hodgman, R.~G. Dall, A.~G. Manning, K.~G.~H. Baldwin, and A.~G. Truscott,
\newblock Science {\bf 331}, 1046 (2011).

\bibitem{Henny:1999cr}
M.~Henny, S.~Oberholzer, C.~Strunk, T.~Heinzel, K.~Ensslin, M.~Holland, and
  C.~Sch\"onenberger,
\newblock Science {\bf 284}, 296 (1999).

\bibitem{Kiesel:2002dq}
H.~Kiesel, A.~Renz, and F.~Hasselbach,
\newblock Nature {\bf 418}, 392 (2002).

\bibitem{Rom:2006uq}
T.~Rom, T.~Best, D.~van Oosten, U.~Schneider, S.~Folling, B.~Paredes, and
  I.~Bloch,
\newblock Nature {\bf 444}, 733 (2006).

\bibitem{Jeltes:2007ly}
T.~Jeltes {\em et~al.},
\newblock Nature {\bf 445}, 402 (2007).

\bibitem{Folling:2005fk}
S.~Folling, F.~Gerbier, A.~Widera, O.~Mandel, T.~Gericke, and I.~Bloch,
\newblock Nature {\bf 434}, 481 (2005).

\bibitem{Hong:1987mz}
C.~K. Hong, Z.~Y. Ou, and L.~Mandel,
\newblock Phys. Rev. Lett. {\bf 59}, 2044 (1987).

\bibitem{PhysRevA.58.4904}
R.~Loudon,
\newblock Phys. Rev. A {\bf 58}, 4904 (1998).

\bibitem{Ou:2006ta}
Z.~Y. Ou,
\newblock Phys. Rev. A {\bf 74}, 063808 (2006).

\bibitem{Ou:1999lo}
Z.~Y. Ou, J.-K. Rhee, and L.~J. Wang,
\newblock Phys. Rev. A {\bf 60}, 593 (1999).

\bibitem{Ou:1999rr}
Z.~Y. Ou, J.-K. Rhee, and L.~J. Wang,
\newblock Phys. Rev. Lett. {\bf 83}, 959 (1999).

\bibitem{Niu:2009pr}
X.-L. Niu, Y.-X. Gong, B.-H. Liu, Y.-F. Huang, G.-C. Guo, and Z.~Y. Ou,
\newblock Optics Letters {\bf 34}, 1297 (2009).

\bibitem{Lim:2005qt}
Y.~L. Lim and A.~Beige,
\newblock N. J. Phys. {\bf 7}, 155 (2005).

\bibitem{Reck:1994zp}
M.~Reck, A.~Zeilinger, H.~J. Bernstein, and P.~Bertani,
\newblock Phys. Rev. Lett. {\bf 73}, 58 (1994).

\bibitem{Weitenberg:2011vn}
C.~Weitenberg, M.~Endres, J.~F. Sherson, M.~Cheneau, P.~Schausz, T.~Fukuhara,
  I.~Bloch, and S.~Kuhr,
\newblock Nature {\bf 471}, 319 (2011).

\bibitem{Tichy:2011vn}
M.~C. Tichy,
\newblock {\em Entanglement and Interference of Identical Particles},
\newblock PhD thesis, University of Freiburg,
  http://www.freidok.uni-freiburg.de/volltexte/8233/, 2011.

\bibitem{Sherson:2010fk}
J.~F. Sherson, C.~Weitenberg, M.~Endres, M.~Cheneau, I.~Bloch, and S.~Kuhr,
\newblock Nature {\bf 467}, 68 (2010).

\bibitem{Bakr:2010fk}
W.~S. Bakr, A.~Peng, M.~E. Tai, R.~Ma, J.~Simon, J.~I. Gillen, S.~F{\"o}lling,
  L.~Pollet, and M.~Greiner,
\newblock Science {\bf 329}, 547 (2010).

\bibitem{Karski:2009vn}
M.~Karski, L.~F{\"o}rster, J.-M. Choi, A.~Steffen, W.~Alt, D.~Meschede, and
  A.~Widera,
\newblock Science {\bf 325}, 174 (2009).

\bibitem{Ryser:1963oa}
H.~J. Ryser,
\newblock {\em Combinatorial Mathematics}~The Carus mathematical monographs
  series (The Mathematical Association of America, 1963).

\bibitem{Sachkov:2002fk}
V.~N. Sachkov and V.~E. Tarakanov,
\newblock {\em Combinatorics of Non-Negative Matrices} (American Mathematical
  Society, 2002).

\bibitem{Minc:1984uq}
H.~Minc and M.~Marcus,
\newblock {\em Permanents} (Cambridge University Press, Cambridge, 1984).

\bibitem{Jerrum:2004:PAA:1008731.1008738}
M.~Jerrum, A.~Sinclair, and E.~Vigoda,
\newblock J. ACM {\bf 51}, 671 (2004).

\bibitem{PhysRevA.55.2564}
M.~\.Zukowski, A.~Zeilinger, and M.~A. Horne,
\newblock Phys. Rev. A {\bf 55}, 2564 (1997).

\bibitem{springerlink:10.1007}
F.~Lalo{\"e} and W.~Mullin,
\newblock Found. Phys. {\bf 42}, 53 (2011).

\bibitem{Graham:1976nx}
R.~L. Graham and D.~H. Lehmer,
\newblock J. Austral. Math. Soc. {\bf 21}, 487 (1976).

\bibitem{Tichy:2010kx}
M.~C. Tichy, M.~Tiersch, F.~de~Melo, F.~Mintert, and A.~Buchleitner,
\newblock Phys. Rev. Lett. {\bf 104}, 220405 (2010).

\bibitem{nockdipm}
M.~Nock,
\newblock Single photons for quantum information processing,
\newblock Master's thesis, Universit\"at Konstanz, 2006.

\bibitem{Campos:1989fk}
R.~A. Campos, B.~E.~A. Saleh, and M.~C. Teich,
\newblock Phys. Rev. A {\bf 40}, 1371 (1989).

\bibitem{al:2006kx}
J.~A. Dunningham, K.~Burnett, R.~Roth, and W.~Phillips,
\newblock N. J. Phys. {\bf 8}, 182 (2006).

\bibitem{al:2006vn}
A.~M. Rey, I.~I. Satija, and C.~W. Clark,
\newblock N. J. Phys. {\bf 8}, 155 (2006).

\bibitem{fourphotons}
M.~Tichy, H.-T. Lim, Y.-S. Ra, F.~Mintert, Y.-H. Kim, and A.~Buchleitner,
\newblock Phys. Rev. A {\bf 83}, 062111 (2011).

\bibitem{Ra:2011ys}
Y.-S. Ra, M.~Tichy, H.-T. Lim, O.~Kwon, F.~Mintert, A.~Buchleitner, and Y.-H.
  Kim,
\newblock Observation of non-monotonic behavior in quantum to classical
  probability transition,
\newblock arXiv:1109.1636, 2011.

\end{thebibliography}
\end{document}